\documentclass[a4paper,oneside,11pt,english]{article}
\usepackage[T1]{fontenc}
\usepackage[latin9]{inputenc}
\usepackage[a4paper]{geometry}
\usepackage{float}
\usepackage{amsmath}
\usepackage{graphicx}
\usepackage{esint}
\usepackage{babel}
\usepackage[margin=35pt,font=footnotesize]{caption}

\usepackage{hyperref}
\hypersetup{colorlinks=true,citecolor=green,linkcolor= blue}
\usepackage{listings}
\usepackage{color}
\usepackage{textcomp}
\definecolor{listinggray}{gray}{0.9}
\definecolor{lbcolor}{rgb}{0.9,0.9,0.9}

\def\sgn{\mbox{sgn}}
\usepackage{bbold}
\def\R{{\mathbb{R}}} 
\def\dotinformula{\;\; \mathrm{.}} 

\begin{document}

\title{\bf Neural Relax}
\author{Elisa Benedetti%
\footnote{Now at: Physics Department T35, Technische Universit\"{a}t M\"{u}nchen, James-Franck Stra\ss e 1, 85747 Garching bei M\"{u}nchen, Germany}%
{\ } and Marco Budinich\\
{\small Physics Department \& INFN, Trieste, Italy}
\\ \\
{\small (Submitted to: \emph{Neural Computation})}
}
\date{ \today }
\maketitle

\abstract{We propose a new self-organizing algorithm for a feed-forward network inspired to an electrostatic problem that turns out to have intimate relations with information maximization.}

\medskip


\section{Introduction}
In this paper we present a new self-organizing algorithm for a layer of $h$ continuous Perceptrons derived from the electrostatic problem of free electrical charges in a conductor. The algorithm is general and maximizes information.

The idea is simple: we use a layer of continuous Perceptrons to map the inputs to point-like electrical charges that we imagine free to move within an hypercube in multi-dimensional space and we let them \emph{evolve}, or better \emph{relax}, under Coulomb repulsion until they set in the minimal energy configuration. For this reason we named this algorithm ``Neural Relax'', NR in what follows.

We show that this is sufficient to obtain binary and statistically independent data as a natural consequence of the algorithm itself, in addition, fixing the dimensions of the hypercube, one can freely adjust the rate of dimensional reduction.
From a theoretical point of view, we show that, in the simple one dimensional case, this algorithm provides the maximum-information solution to the problem, and thus the learning rules result equal to those obtained by Bell and Sejnowski from their Independent Component Analysis (ICA) \cite{Bell_Sejnowski_1995}, exhibiting a completely different interpretation of ICA algorithm. In the general multi-dimensional case, we show that NR gives a pure Hebbian rule and is also well suited to inject some redundancy that can be subsequently used to perform error correction on the processed patterns.

\medskip{}

The paper is structured as follows: in Section \ref{sec:Perceptron_layer} we briefly describe our network. In Section \ref{sec:Analogy} we present the real physical problem we refer to, namely a system of point-like charges confined in a cube, and link it to our problem and to previous works using Coulomb-like forces in neural networks. Then we present a theoretical analysis for the one dimensional case (Section \ref{sec:1D-case}) and the general multi dimensional case (Section \ref{sec:nD case}). We conclude with some preliminary computational results: to test our algorithm we tackle the problem of preprocessing real world binary images to make them unbiased, uncorrelated and binary.

\section{\label{sec:Perceptron_layer}A layer of Perceptrons}
\begin{figure}[h]
\noindent \begin{centering}
\includegraphics[scale=0.15]{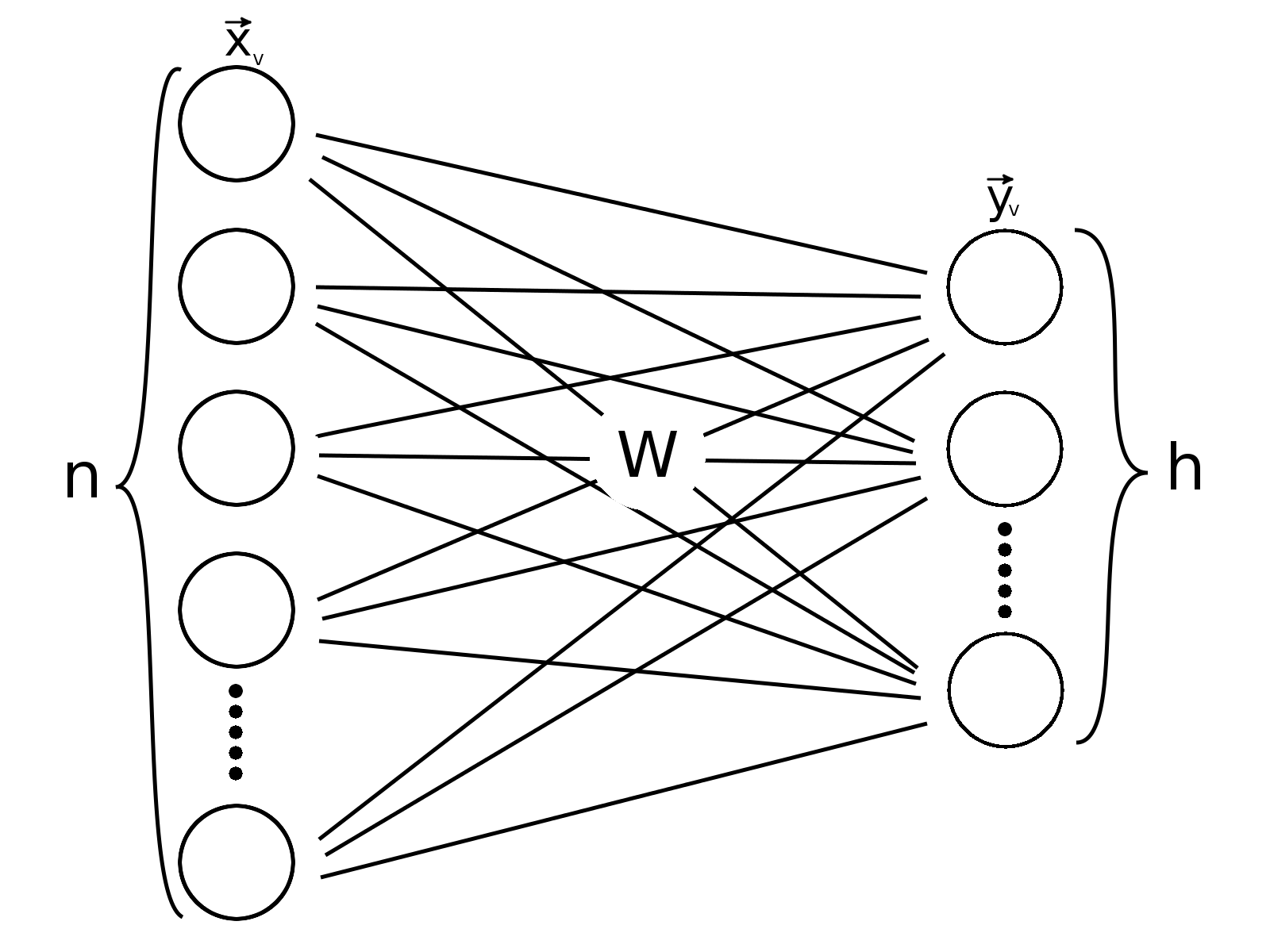}
\par\end{centering}
\caption{Schematic illustration of the network: an input $\vec{x}_\nu$ is fed to an input layer of $n$ neurons, connected to $h$ neurons that produce the output $\vec{y}_\nu$. The weight matrix $W$ contains also the thresholds that appear as weights of a fictitious $0$-th input clamped at $1$.}
\label{fig:topology}
\end{figure}

We consider a layer of $h$ Perceptrons with $n$ inputs and $\tanh()$ transfer function; given an input $\vec{x} \in \R^n$ each Perceptron gives the output
\begin{equation}
\label{eq:y_i_def}
y_{i} = \tanh\left( \vec{w}_i \cdot \vec{x} \right) = \tanh \left( \underset{j = 0}{\overset{n}{\sum}} w_{i j} x_j \right) \qquad i=1,...,h
\end{equation}
and figure~\ref{fig:topology} schematically illustrate the architecture of this network. We stretch a bit the notation indicating the $h$ equations (\ref{eq:y_i_def}) with the weight matrix $W$
\begin{equation}
\label{eq:y_def}
\vec{y} = \tanh \left(W\vec{x}\right) \dotinformula
\end{equation}

This is a common, well studied, network that, among other things, can be used to approximate any continuous function since the transfer function, $\tanh(x)$, is bounded in $(-1,1)$, non constant, smooth and monotone \cite{Hornik_1989}. We will assume that the inputs follow a distribution $p(\vec{x})$ and that there is no noise around; usually we will consider binary inputs $\vec{x} \in \{ \pm 1 \}^n$. We will focus on the case of binary outputs $\vec{y} \in \{ \pm 1 \}^h$ that is the limit of the continuous case (\ref{eq:y_def}) when the argument is large%
\footnote{given that $\lim_{\beta \to \infty} \tanh( \beta x ) = \sgn (x)$}%
.

%

Nadal and Parga \cite{Nadal_Parga_1993} studied this network when $\vec{y} = \sgn \left(W\vec{x}\right)$ in the frame of information theory. They showed that the \emph{information capacity} $C$ that can be conveyed by $h$ binary neurons is bounded by $h$, i.e.
\begin{equation}
C := \underset{p(\vec{x})}{\max} \; I(\vec{x};\vec{y}) \leq h \nonumber
\label{eq:capacity}
\end{equation}
where $I(\vec{x};\vec{y})$ is the mutual information between the input $\vec{x}$, of distribution $p(\vec{x})$, and the output $\vec{y}$. The limitation comes essentially from the architecture since $h$ binary neurons can possibly implement only $C_{h,n}\leq2^{h}$ of the theoretically possible $2^h$ output states and they show that
\begin{equation}
C=\log_{2}C_{h,n}=\begin{cases}
h & \; \mathrm{for} \; h\leq n\\
<h & \; \mathrm{for} \; h>n \dotinformula \end{cases} \nonumber
\end{equation}

So assuming $h \le n$ we see that the architecture doesn't impose any limitation%
\footnote{We just remind that this is different from the request that there is no information loss that depends on the source entropy ${\cal S}(\vec{x})$ and would require that $h \ge S(\vec{x})$.
}%
{} and, for these binary neurons without noise, the upper bound $C$ can be reached if, and only if, the distribution of the outputs $q(\vec{y})$ results fully factorized \cite{Nadal_Parga_1993}, namely
\begin{equation}
q(\vec{y}) = \underset{i=1}{\overset{h}{\prod}}q\left(y_{i}\right) \quad \mbox{with} \quad q\left(y_{i}=\pm1\right)=\frac{1}{2} \quad \forall i \dotinformula
\label{eq:y_independance}
\end{equation}

\medskip

With the help of this analysis we can set up a list of the desirable characteristics for the function $f\,:\, \R^n \rightarrow \R^h$ (\ref{eq:y_def}) implemented by our layer of $h$ Perceptrons:
\begin{itemize}
\item the output patterns should be (essentially) binary i.e.\ $1 - |y_i| < \epsilon$;
\item the map $f\,:\, \R^n \rightarrow \R^h$ should be injective and such that (\ref{eq:y_independance}) holds;
\item as consequence the produced data will be statistically independent:
$$
E[ y_{i_1} y_{i_2} \ldots y_{i_r }] = 0 \quad \forall \; {i_1} \ne {i_2} \ne \cdots \ne {i_r }, \quad \forall \; 1 \le r \le h
$$
(and thus uncorrelated $E[ y_i y_j] = 0 \quad \forall \; i \ne j$);
\item it should accomplish dimensionality reduction i.e.\ whenever possible $h \ll n$;
\item it should be ``learnable'' i.e.\ it should be possible to find it by gradient descent along an appropriate function of the weights.
\end{itemize}

The most demanding goal is satisfying (\ref{eq:y_independance}) but it's not easy to find an algorithm that does it directly. Several authors followed the equivalent path of maximizing the mutual information $I(\vec{x};\vec{y})$, e.g.\ the ICA algorithm \cite{Bell_Sejnowski_1995}; see also \cite{Pham_2001} and references therein. Our algorithm starts from a physical problem that leads naturally towards the fulfillment of these requests.

\section{\label{sec:Analogy}The Physical Problem}
\begin{center}
\begin{figure}[h]
\noindent \begin{centering}
\includegraphics[scale=0.3]{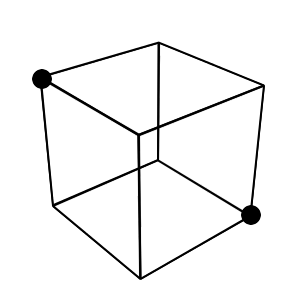}\includegraphics[scale=0.3]{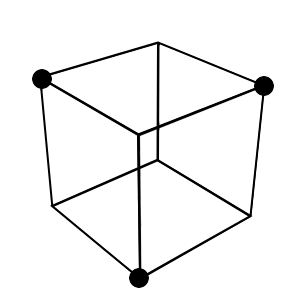}\includegraphics[scale=0.3]{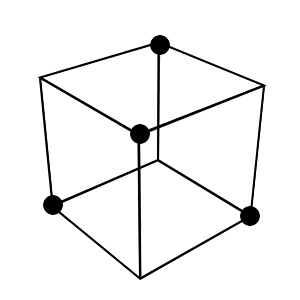}\includegraphics[scale=0.3]{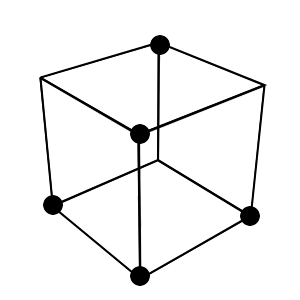}
\par\end{centering}
\caption{{\footnotesize Stable equilibrium configurations of point-like charges in a cubic box: particles arrange in such a way to maximize their reciprocal distances while minimizing the Coulomb potential energy. Since they occupy the vertices they have (almost) binary coordinates in the defined set $H_3$.}}
\label{fig:charges_box}
\end{figure}
\end{center}

Let's consider the problem of finding the stable equilibrium position of $m$, equal, point-like, electric charges $Q_\nu$ within a cube of conductor. This is a problem very similar to the Thomson problem \cite{Thomson_1904} where the charges are in a sphere. Thomson posed it in 1904 and is remarkably difficult to solve, exact solutions are known only for few values of $m$; see \cite{Schwartz_2010}. From now on we will always consider our cube centered at the origin and with side of length $2$, i.e.\ the physical space available to the charges is the $3-$dimensional cube defined by
$$
H_3 = \{ \vec{y} \in \R^3 : |y_i| < 1 \quad i = 1,2,3\}
$$
the extension to the $h$-dimensional hypercube $H_h$ being obvious. In an ideal conductor the $m$ charges are free to move and their stable rest positions $\vec{y}_\nu$ minimize the Coulomb potential%
\footnote{in Gaussian units: $\frac{1}{4 \pi \epsilon_0} = 1$}%
 {} \cite{Jackson_1999}
$$
U \left( \vec{y}_1, \vec{y}_2, \ldots, \vec{y}_m \right) = \underset{\mu<\nu}{\sum}\frac{Q_{\mu} Q_{\nu}}{|\vec{y}_{\mu}-\vec{y}_{\nu}|} \qquad \mu,\nu=1,...,m \dotinformula
\label{eq:Coulomb}
$$
$U \left( \vec{y}_1, \vec{y}_2, \ldots, \vec{y}_m \right)$ is a harmonic function \cite{Axler_2001} and thus doesn't have minima in an open, convex, set like $H_3$, thus the rest positions of the charges are on the border, namely on the surface of the cube. Moreover we \emph{conjecture} that, if the charges are equal and their number is $m \le 2^3 = 8$, the only stable positions of the charges are on cube vertices as shown in Figure~\ref{fig:charges_box}, that contains the minimum energy arrangements for two, three, four and five charges%
\footnote{Despite several attempts we haven't been able to prove this formally but numerical simulations support the conjecture.}%
.

\noindent 

This problem easily generalizes from $\R^3$ to $\R^h$ provided that $U \left( \vec{y}_1, \vec{y}_2, \ldots, \vec{y}_m \right)$ remains harmonic and this happens iff the distance between charges generalizes to
\begin{equation}
\label{eq:generalized_Euclidean_distance}
|\vec{y}_{\mu}-\vec{y}_{\nu}| := [(\vec{y}_{\mu}-\vec{y}_{\nu}) \cdot (\vec{y}_{\mu}-\vec{y}_{\nu})]^\frac{h - 2}{2} \dotinformula
\end{equation}
Also in this case the rest positions of the charges must be on the border of $H_h$ and we generalize our conjecture that charges have stable rest positions on the vertices of $H_h$ and consequently (almost) binary coordinates.

\medskip{}

We take inspiration from this physical problem to propose a self-organizing algorithm for a layer of continuous Perceptrons. We map our set of $m$ inputs in $\R^n$ to point-like charges in $\R^h$ and these charges are bound to remain in the $h$-dimensional hypercube. Subsequently we let this system evolve under Coulomb repulsion in $\R^h$ minimizing its energy until it reaches equilibrium. Provided that our conjecture is true and if $m\leq2^{h}$, the charges at rest will occupy the vertices of $H_h$ and have thus binary coordinates, which means that this approach allows us to get a binary representation of the input data as a natural consequence and without any further constraint. We will also show that this process maximizes information.


More in detail, given a set of $m$ inputs $\vec{x}_\nu \in \R^n, \nu = 1,2, \ldots, m$ of distribution $p(\vec{x})$, applying (\ref{eq:y_def}) we get $m$ outputs $\vec{y}_\nu \in \R^h$ that the hyperbolic tangent constrains within the $h$-dimensional hypercube $H_h$. To treat inputs of different probability $p(\vec{x}_\nu)$ we postulate that the probability of an output $\vec{y}_\nu$ is proportional to the energy of a charge $Q_\nu$ in the electric field, i.e.
\begin{equation}
\label{eq:y_distribution}
q(\vec{y}_\nu) \propto E (Q_\nu) = Q_\nu \underset{\stackrel{\mu = 1}{\mu \ne \nu}}{\overset{m}{\sum}} \frac{Q_\mu}{|\vec{y}_{\mu}-\vec{y}_{\nu}|}
\end{equation}
and the total energy of the system is:
\begin{equation}
\label{eq:potential}
U \left( \vec{y}_1, \vec{y}_2, \ldots, \vec{y}_m \right) = \sum_{\nu = 1}^m E (Q_\nu) = \sum_{\mu<\nu} \frac{Q_\mu Q_\nu}{|\vec{y}_{\mu}-\vec{y}_{\nu}|} \dotinformula
\end{equation}
For the sake of simplicity most of the times we will assume that all inputs are equiprobable $p(\vec{x}_\nu) = \frac{1}{m}$ and thus we will feel free to put $Q_\nu = 1$ for all $m$ charges and the function to minimize is the simplified Coulomb potential
\begin{equation}
\label{eq:simplified_potential}
U \left( \vec{y}_1, \vec{y}_2, \ldots, \vec{y}_m \right) = \underset{\mu<\nu}{\sum}\frac{1}{|\vec{y}_{\mu}-\vec{y}_{\nu}|} \qquad \mu,\nu=1,...,m \dotinformula
\end{equation}
This ``energy'' is the function that NR learning algorithm minimizes modifying the elements of the weight matrix $W$ by gradient descent namely
\begin{equation}
\label{eq:gradient_descent}
w_{i j}' = w_{i j}-\epsilon\frac{\partial U \left( \vec{y}_1, \vec{y}_2, \ldots, \vec{y}_m \right)}{\partial w_{i j}}
\end{equation}
$\epsilon$ being a small positive constant.

Let us suppose that NR has been successfully applied and that the harmonic function $U$ has been minimized (more on this later). All the $m$ charges have relaxed in the minimum energy configuration and necessarily lie on $H_h$ surface and, if $m \le 2^h$ and our conjecture is true, they sit precisely on the vertices of the hypercube $H_h$. It follows that all coordinates of their positions $\vec{y}_\nu$ are binary and represent satisfactorily the outputs of $h$ binary neurons.

With distance definition (\ref{eq:generalized_Euclidean_distance}) we know that $U$ is harmonic and Gauss theorem holds. We use these properties to show that the positions of our charges satisfy (\ref{eq:y_independance}) in the limit $n, m, h \to \infty$ when we can neglect the granularity of the charges and we can assume that the charge distribution becomes continuous. A similar approach is usually taken for idealized physical conductors where one forgets the quantization of electron charges since the single electron charge is considered negligible with respect to the total charge on the conductor.

When the charges have relaxed in the minimum energy configuration we know that there is no electric field within the conductors and that all charges lie on the (hyper-)surface, moreover the spatial density of the charges must be constant in the limit $n, m, h \to \infty$. It follows, given the $H_h$ structure%
\footnote{one can observe that if the charges sit on hypercube vertices they also lie on the hypersphere of radius $h^{\frac{h-2}{2}}$ and continue the following proofs for the hypersphere}%
, that every hyperplane through the origin of $\R^h$ and that doesn't hit any vertex of $H_h$ (to avoid complications) cuts $H_h$ into two parts that contain the same number of vertices, since, if vertex $\vec{v}$ belongs to one of the semi-spaces, vertex $-\vec{v}$ must belong to the other one. From the constancy of the spatial density of the charges it follows that the two semi-spaces must also contain exactly the same charge, one half of the total charge on $H_h$. Since this results is valid for any hyperplane through the origin of $\R^h$ it is true also for the $h$ hyperplanes $y_i = 0$. This means that there are exactly $\frac{m}{2}$ charges with $y_i = 1$ (remember all coordinates are binary) and the same number with $y_i = -1$. In the language of our layer of Perceptrons and since $m \to \infty$ this means that the output distribution is such that
$$
q\left(y_{i}=\pm1\right)=\frac{1}{2} \quad \forall i \dotinformula
$$
It's also easy to prove by induction that $q(\vec{y}) = \underset{i=1}{\overset{h}{\prod}}q\left(y_{i}\right)$, we begin showing that $q(y_{i}, y_{j})=q(y_{i}) q(y_{j})$ for any couple of different coordinates $y_i$ and $y_j$. Let's suppose we have cut our charge distribution into two equal parts by the hyperplane $y_i = 0$ and we consider the orthogonal hyperplane $y_j= 0$, it's easy to use the previous argument to show that in all 4 subspaces so defined the charges must be equal to $\frac{m}{4}$ and thus that for any choices of the values of $y_i$ and $y_j$ one gets $q(y_{i}, y_{j})=\frac{1}{4}$ and thus $q(y_{i}, y_{j})=q(y_{i}) q(y_{j})$. Let's now suppose $q(y_{i_1}, y_{i_2}, \dots, y_{i_k}) = \underset{j=1}{\overset{k}{\prod}}q\left(y_{i_j}\right) = \frac{1}{2^k}$ for any choice of $k$ variables $y_{i_1}, y_{i_2}, \dots, y_{i_k}$, it's easy to exploit the structure of $H_h$ to show that, if one adds a $(k+1)$-th coordinate, the hyperplane of equation $y_{i_{k+1}} = 0$ will cut all the previous charges into 2 halves and thus that $q(y_{i_1}, y_{i_2}, \dots, y_{i_k}, y_{i_{k+1}}) = \underset{j=1}{\overset{k+1}{\prod}}q\left(y_{i_j}\right) = \frac{1}{2^{k+1}}$ completing the proof by induction. A technical point: we note that only for $m = 2^h$ one can continue the induction chain up to step $k = h$ giving $q(\vec{y}) = 2^{-h}$ for any $\vec{y}$ and complete factorization of the distribution $q(\vec{y})$; if $m < 2^h$ one can only prove that all the moments of order $k$ of $q(\vec{y})$ are zero up to $k = \lfloor \log_2 m \rfloor$.

We have thus proved that, if the $m$ charges relax in the configuration of minimal energy (that by the way it's far from being unique given the many symmetries of the system) the final positions of the charges satisfy all the requests set for a layer of Perceptrons at the end of the previous Section, in particular that the final distribution is fully factorized (\ref{eq:y_independance}) that implies that the information produced at the output is maximal.

\medskip

There is one point we left behind that deserves attention: we saw that $U \left( \vec{y}_1, \vec{y}_2, \ldots, \vec{y}_m \right)$ is a function of the $m h$ charges coordinates and is provably harmonic, but in our case, with (\ref{eq:y_i_def}), we can change $\vec{y}_i$ coordinates only through the $(n+1) h$ weights $w_{i j}$. It is simple to verify that $U(w_{i j})$ is no more harmonic:
\begin{eqnarray*}
\frac{\partial U}{\partial w_{i j}} & = & \frac{\partial U}{\partial y_{i}} \frac{\partial y_{i}}{\partial w_{i j}}\\
\frac{\partial^2 U}{\partial w_{i j}^2} & = & \frac{\partial^2 U}{\partial y_{i}^2} \left( \frac{\partial y_{i}}{\partial w_{i j}} \right)^2 + \frac{\partial U}{\partial y_{i}} \frac{\partial^2 y_{i}}{\partial w_{i j}^2}
\end{eqnarray*}
and in general $\nabla^2 U(w_{i j}) = \sum_{i,j} \frac{\partial^2 U}{\partial w_{i j}^2} \neq 0$. This means that the restrictions imposed to the positions of the charges $\vec{y}_\nu$ by the fact that they are defined by $\vec{y} = \tanh\left(W\vec{x}\right)$ --- that, by the way, enforces also the constraints $\vec{y}_\nu \in H_h$ --- renders the energy no more harmonic in the ``free'' coordinates $w_{i j}$. This implies that we cannot formally {\em prove} that the function $U(w_{i j})$ is without local minima and that gradient descent (\ref{eq:gradient_descent}) will always bring the system to one of the solutions we just described essentially because we cannot ``move'' freely the charge positions $\vec{y}_\nu$ but only through the variation of the weights $w_{i j}$.

One could argue that it is reasonable to expect that the characteristics of the found solution won't change dramatically, especially if $m \ll 2^h$, and the charges are very far from each other on $H_h$, but still the strength of a formal proof is lost. This argument surely deserves further investigations and will be the subject of future work.

\medskip
 
We conclude this section with a brief review of other appearances of Coulomb-like forces in the context of neural networks. The series started in 1987 when Bachmann et al. \cite{Bachmann_Cooper_1987} proposed an associative memory that attached negative electrical charges to the stored patterns and the memory played the role of a positive charge attracted by the patterns. In this fashion they could store unlimited patterns and the memory didn't have any spurious state. This idea resurged 7 year later \cite{Perrone_Cooper_1995}.

After some years Marques and Almeida \cite{Marques_Almeida_1999} proposed a feed forward network dedicated to the separation of nonlinear mixtures that minimized a function made of three terms. The first term, $W$, was inspired to the idea of repulsion of equal charges and produced a repulsive force. This force was non-physical since the repulsion had a finite range and acted only in proximity of the patterns; the minimization of this term tended to keep the patterns far apart producing an approximately uniform distribution of the patterns. To this term they had to add a term $B$, enforcing the constraints of the outputs in $[-1,1]$ not to have the patterns fly to infinity and a regularizing term $R$. This work has been subsequently analyzed in a mathematical setting \cite{Theis_et_al_2001} where it has been shown that, within certain approximations, a repulsive force decreasing faster than the Coulomb force, tends to produce uniform probability density of the outputs that in turn maximizes output entropy that in turn minimizes mutual information and is thus amenable to ICA.

All of these works do not have a real, physical, Coulomb energy that is instead central in our approach since it will allow us to define properly a positive definite probability density (\ref{eq:pot_en1}) and will provide an energy that, at least in the ideal case, is harmonic and thus gives important properties to the function to be minimized. This kind of potential matches perfectly with the hypercube structure since charges tend to put themselves on the hypercube vertices thus automatically satisfying the other request of having binary coordinates. This produces a distribution of the patterns that, microscopically, is highly non uniform, being the discrete sum of point-like charges. On the other hand, from a larger distance, this distribution appears uniform thanks to Gauss theorem (as happens in real conductors).

\section{\label{sec:1D-case}Analysis of the 1-dimensional Case}
We start analyzing NR properties in a toy problem: a layer made of just one neuron with one input; i.e.\ a purely one dimensional problem. This is a well studied case \cite{Atick_1992, Nadal_Parga_1994, Bell_Sejnowski_1995} where theoretical analysis is simpler: here (\ref{eq:y_i_def}) becomes
\begin{equation}
y = \tanh (w x + w_0) \dotinformula
\label{eq:y_one_dim}
\end{equation}

Only for the analysis of this case we relax the condition of digital inputs since this would restrict us to the too simple case $x = \pm1$. So here we suppose to have continuos inputs $x$ with probability distribution $p(x)$. Correspondingly we have continuos $y$ with an electrical charge density $\rho(y)$ and the energy of the system (\ref{eq:potential}) becomes:
$$
U = \iint \! \frac{\rho\left(y\right)\rho\left(y'\right)}{|y-y'|} \, \mathrm{d}y\mathrm{d}y'
$$
calling $\phi\left(y\right) := \int \! \frac{\rho\left(y'\right)}{|y-y'|} \, \mathrm{d}y'$ the total potential of point $y$, we have
\begin{equation}
U=\int \! \rho\left(y\right) \phi\left(y\right) \, \mathrm{d}y := \int \! q(y) \, \mathrm{d}y
\label{eq:pot_en1}
\end{equation}
where $q(y)$ is the linear energy density that is by definition positive since it is proportional to the squared electric field \cite{Jackson_1999}. It is thus possible to extend (\ref{eq:y_distribution}) and to interpret $q(y)$ (suitably normalized) also as the probability density distribution of $y$. Our problem is, given $x$ and $p(x)$, to determine the parameters $w, w_0$ that minimize $U$.

We can gain insight into the actual solution of this problem examining first the corresponding physical problem: since our charges in $y$ are to be imagined as free charges in a conductor this is the physical problem of the charge distribution on a finite (remember $-1 < y < 1$), infinitely thin, conductive wire.

It is a typical electrostatic problem: one has to find the charge distribution $\rho\left(y\right)$ that minimizes $U$. In this particular case we are in a conductor and thus, when the energy is minimized, the potential is constant $\phi\left(y\right) = \phi_0$ and so mathematically the problem is to find the charge distribution $\rho\left(y\right)$ that realizes this condition. This is not an easy problem (it has been the subject of James Clerk Maxwell's last scientific paper, see in \cite{Jackson_2002}) but is known \cite{Jackson_2000} that, as the ratio of the physical dimensions of the wire goes to zero, the distribution of the charges on the wire $\rho\left(y\right)$ tends to a uniform distribution, i.e.\ $\rho\left(y\right) \to \rho_0$. So we can conclude that the physical solution that minimizes \eqref{eq:pot_en1} gives $q(y) = \rho_0 \phi_0$.

This is true for the physical problem where, since the charges in the wire are free to move, the distribution of charges $\rho\left(y\right)$ can take any shape. Viceversa it is clear that in our case, where we can play only with the parameters $w, w_0$ to modify $\rho\left(y\right)$, in general it will be impossible to find values of $w, w_0$ that realize the condition $q(y) = \rho_0 \phi_0$.

But let us suppose that we are in this lucky case; to understand what is the meaning for our problem we use the well known relation for the transformation of a distribution $p(x)$ when the variable $x$ is transformed to $y = f_w(x)$ where $w$ represent the parameters of the function $f()$ that has to be invertible. In this case the distribution $q(y)$ of $y$ is given by
$$
q(y) = \frac{p(x)}{|\frac{\partial f_w(x)}{\partial x}|}
$$
and this relation tells us that to get a constant $q(y)$ necessarily $|\frac{\partial f_w(x)}{\partial x}| \propto p(x)$ and thus the function $y = f_w(x)$ needs to be proportional to the primitive of the probability distribution of $x$, namely
\begin{equation}
\label{eq:max_info_sol}
f_w(x) \propto \int \! p(x) \, \mathrm{d}x
\end{equation}
and it's well known that this represents the maximum entropy solution for our one-neuron net \cite{Atick_1992}. So, if adjusting $w$ and $w_0$ we can obtain that indeed (\ref{eq:max_info_sol}) holds, our system minimizes energy (\ref{eq:pot_en1}) and this solution gives also the maximum information. In our case (\ref{eq:y_one_dim}) one obtains:
$$
\tanh'(wx+w_{0}) |w| \propto p(x)
$$
where we used the fact that $\tanh'(x) > 0$, and this relation can also be interpreted to give the only possible $p(x)$ for which we get the optimal solution. As pointed out by one of the referees this can be a severe limitation to which one could put remedy adapting not just the weights but, as done in \cite{Nadal_Parga_1994}, the transfer function itself $f_w(x)$. This would produce a more powerful neuron but, following Bell and Sejnowski's ICA, we decided purposely not to open this Pandora's jar at this stage.

Now we analyze what happens in the general case when (\ref{eq:max_info_sol}) can't be satisfied exactly and the best one can do is to find the values of the parameters $w$ that minimize $U$, i.e.\ we study
\begin{equation}
\label{eq:energy_gradient}
\frac{\partial U}{\partial w} = \frac{\partial}{\partial w} \int \! q(y) \, \mathrm{d}y = \int \! \frac{\partial}{\partial w} \frac{p(x)}{|\frac{\partial f_w(x)}{\partial x}|} \, \mathrm{d}x
\end{equation}
where we applied Leibnitz's rule for differentiation under the integral since we are dealing with continuous functions. We observe that the only term that depends on $w$, and is thus affected by the derivative, is $| \frac{\partial f_w(x)}{\partial x} |$.

We conclude this Section showing that the learning rules for our network (\ref{eq:y_one_dim}), obtained by (\ref{eq:energy_gradient}), are equivalent to the Bell and Sejnowski's ICA \cite{Bell_Sejnowski_1995}. We start performing the derivation with respect to $w$ and $w_{0}$
\begin{eqnarray*}
- \frac{\partial U}{\partial w} & = & \int \! \frac{p(x)}{\left[f_w\,'\left(wx+w_{0}\right)|w|\right]^{2}} \left[f_w\,''\left(wx+w_{0}\right)|w|x+\frac{|w|}{w} f_w\,'\left(wx+w_{0}\right)\right] \, \mathrm{d}x \nonumber \\
- \frac{\partial U}{\partial w_{0}} & = & \int \! \frac{p(x)}{\left[f_w\,'\left(wx+w_{0}\right)|w|\right]^{2}} \left[f_w\,''\left(wx+w_{0}\right)|w|\right] \, \mathrm{d}x
\label{eq:deriv_Uw0}
\end{eqnarray*}
with our choice $y=f_w\left(wx+w_{0}\right)=\tanh\left(wx+w_{0}\right)$; then
\begin{equation}
\begin{cases}
f_w\,'\left(wx+w_{0}\right)=1-y^{2} > 0\\
f_w\,''\left(wx+w_{0}\right)=-2y\left(1-y^{2}\right)\end{cases}
\label{eq:deriv_tanh}\end{equation}
that substituted in previous equations give
\begin{eqnarray*}
- \frac{\partial U}{\partial w} & = & \int \! \frac{p(x)}{\left[\left(1-y^{2}\right)|w|\right]^{2}} \left[-2y\left(1-y^{2}\right)|w|x+\frac{|w|}{w}\left(1-y^{2}\right)\right] \, \mathrm{d}x=\nonumber \\
 & = & \int \! \frac{p(x)}{\left(1-y^{2}\right)|w|^{2}} \left[-2y|w|x+\frac{|w|}{w}\right] \, \mathrm{d}x = \nonumber \\
 & = & \int \! \frac{p(x)}{\left(1-y^{2}\right)|w|} \left[\frac{1}{w}-2yx\right] \, \mathrm{d}x \nonumber \\
- \frac{\partial U}{\partial w_{0}} & = & \int \! \frac{p(x)}{\left[\left(1-y^{2}\right)|w|\right]^{2}} \left[-2y\left(1-y^{2}\right)|w|\right] \, \mathrm{d}x=\nonumber \\
 & = & \int \! \frac{p(x)}{\left(1-y^{2}\right)|w|^{2}} \left[-2y|w|\right] \, \mathrm{d}x = \int \! \frac{p(x)}{\left(1-y^{2}\right)|w|} \left[-2y\right] \, \mathrm{d}x \dotinformula
\end{eqnarray*}

Comparing these equations with (\ref{eq:pot_en1}) we note that the term $\int \! \frac{p(x)}{\left(1-y^{2}\right)|w|} \, \mathrm{d}x$ is nothing but the Coulomb energy $U$ integrated over $x$, and hence, as anticipated, it is possible to interpret it as a distribution over which the terms in square brackets can be considered \emph{averaged}, so we can also write them as expectation values:
\begin{equation}
\begin{cases}
- \frac{\partial U}{\partial w} & = \; E_{U}\left[\frac{1}{w}-2yx\right]\\
- \frac{\partial U}{\partial w_{0}} & = \; E_{U}\left[-2y\right]
\end{cases} \nonumber
\end{equation}
and comparing these relations with ICA's learning rules \cite{Bell_Sejnowski_1995} (remembering that we use slightly different transfer functions), we see that they are equal. This shows that NR and ICA are intimately related and that, even if they start from completely different starting points, essentially they both end up maximizing information.

\section{\label{sec:nD case}The Multidimensional Case}
We now proceed to examine the general multidimensional case: we start with $m$ binary inputs of $n$ bits each (that in our numerical simulations will be binary images)
$$
\vec{x}_\nu \in \{ \pm 1 \}^n \qquad \nu = 1, 2, \ldots, m
$$
fed to a layer of $h$ neurons thus producing, for each input,
$$
\vec{y}_\nu = \tanh( W \vec{x}_\nu ) \in (-1,1)^h \qquad \nu = 1, 2, \ldots, m
$$
where the dimensionality of the output layer $h$ is a quite arbitrary choice: it represents somehow the compression rate of the system%
\footnote{as proposed in \cite{Nadal_Parga_1993} one can distinguish $3$ cases:
\begin{itemize}
\item[-] $h < {\cal S}(\vec{x})$ here the net must ``compress'' the data with some information loss;
\item[-] $h = {\cal S}(\vec{x})$ here the net is perfectly matched to the incoming information;
\item[-] $h > {\cal S}(\vec{x})$ here the net is redundant but, as explained later, with NR this redundancy can be used for error correction.
\end{itemize}
}%
. To each output $\vec{y}_\nu$ produced we attach an arbitrary unitary electric charge. Then we calculate the Coulomb potential (\ref{eq:simplified_potential}) and apply gradient descent to it to obtain the learning rules. With the standard distance definition (\ref{eq:generalized_Euclidean_distance}) in $h-$dimensional space we get
$$
|\vec{y}_{\mu}-\vec{y}_{\nu}| = \left[ \underset{i=1}{\overset{h}{\sum}}\left(y_{\mu i}-y_{\nu i}\right)^{2} \right]^\frac{h - 2}{2}
$$
that gives the learning rule for $h > 2$
\begin{eqnarray}
\label{eq:gradient_U}
\Delta w_{ij} & = & -\frac{\partial U}{\partial w_{ij}} = -\frac{\partial}{\partial w_{ij}}\underset{\mu<\nu}{\sum}\frac{1}{|\vec{y}_{\mu}-\vec{y}_{\nu}|} = \nonumber \\
& = & \underset{\mu<\nu}{\sum} \frac{2 - h}{|\vec{y}_{\mu}-\vec{y}_{\nu}|^{\frac{h}{h-2}}}\left(y_{\nu i}-y_{\mu i}\right)\left[x_{\mu j}\left(1-y_{\mu i}^{2}\right)-x_{\nu j}\left(1-y_{\nu i}^{2}\right)\right]
\end{eqnarray}
where we used the properties (\ref{eq:deriv_tanh}) of the hyperbolic tangent.

We used the only possible definition of the distance $|\vec{y}_{\mu}-\vec{y}_{\nu}|$ that renders the energy $U$ harmonic in the $m h$ variables $y_{\nu i}$ but this is of little use for us since in general $U$ is not harmonic with respect to our ``free'' variables $w_{i j}$.

We have thus felt free to try another definition for the distance with the objective of obtaining a faster learning algorithm.
For these reasons we considered the expression

\begin{equation}
|\vec{y}_{\mu}-\vec{y}_{\nu}|_H := \left[2 \left(h - \vec{y}_{\mu} \cdot \vec{y}_{\nu} \right) \right]^\frac{h - 2}{2} = \left[ 2h \left(1-\frac{\underset{i=1}{\overset{h}{\sum}} y_{\mu i} y_{\nu i}}{h}\right)\right]^\frac{h - 2}{2}\nonumber
\end{equation}
that is a distance in mathematical sense; indeed it is a slightly modified version of the so called Hamming distance, which is a measure of the difference between two strings of equal length%
\footnote{in our notation the Hamming distance between binary vectors $\vec{y}_{\mu}, \vec{y}_{\nu} \in \{ \pm 1\}^h$ is $\frac{1}{2} (h - \vec{y}_{\mu} \cdot \vec{y}_{\nu})$}%
. With this new distance plugged in (\ref{eq:simplified_potential}) we define a slightly different energy function $U_H$ that still diverges when any two charges get too near to each other. Minimizing $U_H$ the learning rule becomes
\begin{eqnarray}
\Delta w_{ij} & = & -\frac{\partial U_H}{\partial w_{ij}} = -\frac{\partial}{\partial w_{ij}} \underset{\mu<\nu}{\sum} \; \frac{1}{|\vec{y}_{\mu}-\vec{y}_{\nu}|_H}= \nonumber \\
\label{eq:gradient_U_H}
 & = & \underset{\mu<\nu}{\sum} \; \frac{2 - h}{|\vec{y}_{\mu}-\vec{y}_{\nu}|_H^{\frac{h}{h-2}}} \; \left[x_{\mu j}y_{\nu i} (1 - y_{\mu i}^2)+x_{\nu j}y_{\mu i} (1-y_{\nu i}^2) \right]
\end{eqnarray}
that is similar to previous rule (\ref{eq:gradient_U}) with the only difference that it contains only the ``crossed'' Hebbian terms $x_{\mu j}y_{\nu i}$ and $x_{\nu j}y_{\mu i}$ without the subtraction of the ``straight'' terms $x_{\mu j} y_{\mu i}$ and $x_{\nu j} y_{\nu i}$ and that in numerical simulation appears indeed to be faster.

This modified Hamming distance can be easily related to the Euclidean distance (\ref{eq:generalized_Euclidean_distance}) observing that since the output of the hyperbolic tangent is in $( -1, 1)$ it follows that $0 \le \vec{y}^{2} \le h$ and so
\begin{eqnarray}
|\vec{y}_{\mu}-\vec{y}_{\nu}| & = & \left[\left(\vec{y}_{\mu}-\vec{y}_{\nu}\right) \cdot \left(\vec{y}_{\mu}-\vec{y}_{\nu}\right) \right]^\frac{h - 2}{2} = \left[\vec{y}_{\mu}^{2}+\vec{y}_{\nu}^{2}-2\vec{y}_{\mu} \cdot \vec{y}_{\nu} \right]^\frac{h - 2}{2} \nonumber \\
 & \leq & \left[ 2 (h - \vec{y}_{\mu} \cdot \vec{y}_{\nu}) \right]^\frac{h - 2}{2} = |\vec{y}_{\mu}-\vec{y}_{\nu}|_H \qquad \qquad \forall \vec{y} \in H_h \nonumber
\label{eq:distances_inequality}
\end{eqnarray}
and the Euclidean and the Hamming distances coincide if, and only if, each component of each output vector is binary, which is basically what we hope to get at equilibrium. In terms of the energy we can thus write
\begin{equation}
U \left( \vec{y}_1, \vec{y}_2, \ldots, \vec{y}_m \right) \geq U_H \left( \vec{y}_1, \vec{y}_2, \ldots, \vec{y}_m \right) \qquad \qquad \forall \vec{y} \in H_h
\label{eq:U_distances}
\end{equation}
and we see then that the energy defined with the Hamming distance is a lower bound for the energy defined making use of the Euclidean one. In principle, thus, at equilibrium we can expect the two energies to be equal.

\smallskip

Learning rules (\ref{eq:gradient_U}) and (\ref{eq:gradient_U_H}) share two characteristics: the first is that they are Hebbian since they are perfectly local in the sense that the synapse $w_{i j}$ connecting neuron $y_i$ to input $x_j$ is updated only with the values taken by these neurons. At the same time the value of the synapse is updated by the product $x_j y_i$ referring only to different patterns: in other words to update a synapse one needs the ``history'' of the two neurons; one could say that the rule is local in space but non-local in time. The second interesting characteristic is that in both rules appear the terms $(1-y_{\nu i}^{2})$ that tend to kill the learning when $|y_{\nu i}| \simeq 1$, i.e.\ when the coordinates are substantially binary; this inhibits the weights from growing indefinitely.

\bigskip

We conclude this Section observing that the outputs produced by this network are suited to implement error detection and correction, in other words the injective map $f\,:\, \R^n \rightarrow \R^h$ (\ref{eq:y_def}) implemented by our network de facto acts as an encoder that realizes a block $(m,h)$ code, see e.g.\ \cite{Cover_Thomas}. Let's suppose that $m < 2^h$, i.e. there are less patterns $\vec{y}_\nu$ then hypercube vertices to park them and that $U$ has been minimized. Given the form of the energy minimized by learning (\ref{eq:simplified_potential}) we know that each charge $\vec{y}_\nu$ will be on a hypercube vertex and as far as possible from all other charges. Let us suppose that the minimum Hamming distance between different charges $\vec{y}_\nu$ is $d$, it's well known that in this case one can detect up to $d-1$ errors on the patterns $\vec{y}$ and correct up to $\lfloor \frac{d-1}{2} \rfloor$ errors. For example in the numerical simulations of the next Section, for $m=7$ and $h=64$, the minimum Hamming distance between different patterns is larger than $d = 36$.

This means that if one is given a noisy version $\vec{y}'_\nu$ of the pattern $\vec{y}_\nu$ (for example as returned by an associative memory) one can try to restore the original pattern. By the way the restoration could be done by minimizing again the potential energy $U$ that it's no more minimal when the correct pattern $\vec{y}_\nu$ is replaced by its noisy version $\vec{y}'_\nu$ that results ``out of place''.

\section{\label{sec:Results}Preliminary Numerical Results}
We start introducing the problem we tackled to test NR namely the preprocessing of real world data to build a binary, uncorrelated representation. We had in mind the preprocessing of binary images for an associative memory but this task is by no means limited to this particular problem.

Associative memories have been one of the first applications of the neural networks paradigm: introduced in 1969 by David Willshaw et al. \cite{Willshaw_1969} have produced many offsprings: see e.g.\ the classical book \cite{Hertz_Krogh_Palmer_1991} and references therein, or, for a more recent review, see \cite{Knoblauch_2011} that embeds all flavours of associative memories in a unique Bayesian frame. We focus on the (classical) family of associative memories made of a network of $n$ McCulloch and Pitts neurons each of them updating its state $S_i \rightarrow S_i^\prime$ with the standard rule
\begin{equation}
S_i^\prime = t\left(\sum_{j=1}^n w_{i j} S_j \right)
\label{eq:update}
\end{equation}
where the transfer function $t(x)$ can be either smooth, e.g.\ $t(x) = \tanh(x)$, or binary, $t(x) = \sgn(x)$. Different kinds of associative memories sport different connection schemes and different rules for the synapses $w_{i j}$ but all models agree on the fact that the information is stored in synapses. An associative memory storing $m$ patterns $\vec{\xi_{\nu}} ,\; \nu = 1,\ldots,m$ should be able to find any of the stored patterns starting from a partial or noisy cue. More precisely if the network is initially in state $\vec{S}_0$ the (repeated) application of (\ref{eq:update}) should bring the network in one of the stored states i.e.\ $\vec{S}_0 \rightarrow \vec{S} = \vec{\xi_{\nu}}$.

A common simplification easing analytical calculations is that of assuming the distribution of the stored patterns to be fully factorized and unbiased:
\begin{equation}
P\left(\vec{\xi}\right)=\underset{i=1}{\overset{n}{\prod}}p\left(\xi_{i}\right) \quad \mbox{with} \quad p\left(\xi_{i}=\pm1\right)=\frac{1}{2} \quad \forall i
\label{eq:factorization}
\end{equation}
that implies that the patterns are statistically independent and binary. This request is exacting and, if it's strictly respected, rules out immediately all real world data like for example binary images or sparse coded data.

So to deal with these data one needs to transform them first in data that fulfills these requirements. The simplest transformations are the linear ones and if one contents himself with uncorrelated data (and not independent) than the linear transformation known as Principal Component Analysis can do the job. Unfortunately the transformed patterns are no more binary and it is an open problem to find a linear transformation that produces uncorrelated {\em and} binary data (see e.g.\ \cite{Tang_2006} or \cite{Schein_2003}, an exact solution being in general impossible%
\footnote{the covariance matrix has integer elements but this is not true for its eigenvectors.}%
). So to end up with binary data one must give up to one of the constraints: uncorrelation or linearity of the transformation.

Here we abandon the request of a linear transformation but doing that we can at the same time soar our other goal: we will produce data that is not just uncorrelated but independent, while at the same time remaining binary. More precisely, given $m$, $n$-dimensional, binary images $\vec{\xi_{\nu}}$, we look for
$$
f\,:\, \R^n \rightarrow \R^h \qquad \vec{y}_{\nu} = f(\vec{\xi_{\nu}}) \qquad \mbox{with} \quad h \le n
$$
and the outputs $\vec{y}_{\nu}$ represent the preprocessed patterns that should be statistically independent and thus ready to be stored in an associative memory of $h$ neurons. At this point it's clear that (\ref{eq:y_def}) obtained by NR, that satisfies (\ref{eq:y_independance}), it's tailored for the job.

Before presenting numerical results we just mention an additional complication due to the fact that associative memories usually do not recall exactly the stored patterns $\vec{y}_{\nu}$ but return the pattern $\vec{S} = \vec{y}_{\nu}^\prime$ with $\vec{y}_{\nu}^\prime \simeq \vec{y}_{\nu}$ the difference being typically a few percent of the bits. If one wants to be able to get back the original image $\vec{\xi_{\nu}}$ from $\vec{y}_{\nu}^\prime$ this imposes further requirements to the characteristics of the preprocessing while, at the same time, rules out standard algorithms for binary compression that produce statistically fragile data. As explained in previous Section NR, providing data that are as much farther apart as possible in $R^h$, can fulfill also this request.

\medskip

We run a preliminary numerical test on a set of $m = 7$ binary images of $33\times33$ pixels; we had a network of $h = 64$ neurons with $n=33\times33+1=1,\!090$ inputs totalling $69,\!760$ weights. We run two different learning runs with the two gradient descent rules (\ref{eq:gradient_U}) and (\ref{eq:gradient_U_H}) of previous Section. The program stopped when $\underset{i,j}{\max}\left\{ \Delta w_{ij}\right\} \leq10^{-5}$ that required of the order of $10^{7}$ steps. Each simulation took several days of an Intel Core Duo 2.93 GHz processor indicating that there is ample space for improvements, e.g.\ by taking advantage from standard electrostatics relaxing algorithms.

\begin{figure}[h]
\noindent \begin{centering}
\includegraphics[scale=0.4]{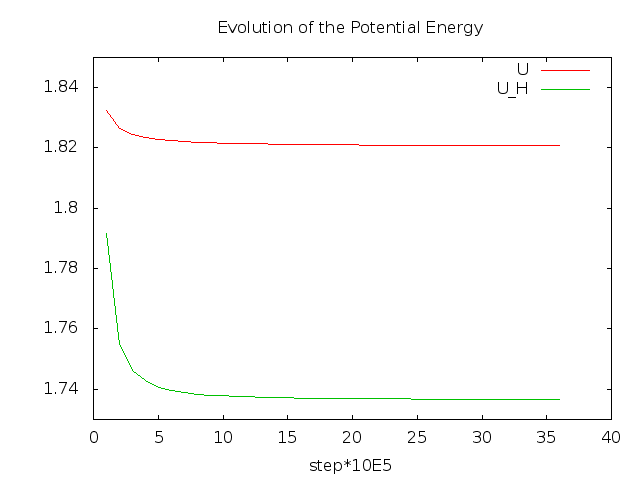}
\par\end{centering}
\noindent\caption{{\footnotesize Behaviour of the energy during the run for both the Euclidean, $U$, and the Hamming distance $U_H$: as can be seen, the latter is smaller than the first, as predicted by (\ref{eq:U_distances}). The x axis represent the running step in unit of $10^{5}$ elementary steps.}}
\label{fig:potential_energies}
\end{figure}

Figure~\ref{fig:potential_energies} show the energy decrease during learning for both the Euclidean $U$ and the Hamming distance $U_H$: the first impression is that, as one could expect, the decrease is compatible with a typical electrostatic potential; also $U \ge U_H$ as foreseen. In this first run the expected convergence of $U \to U_H$ was not observed but there are indications that $U$ minimization was not terminated.

%

\begin{figure}[H]
\noindent \begin{centering}
\includegraphics[scale=0.5]{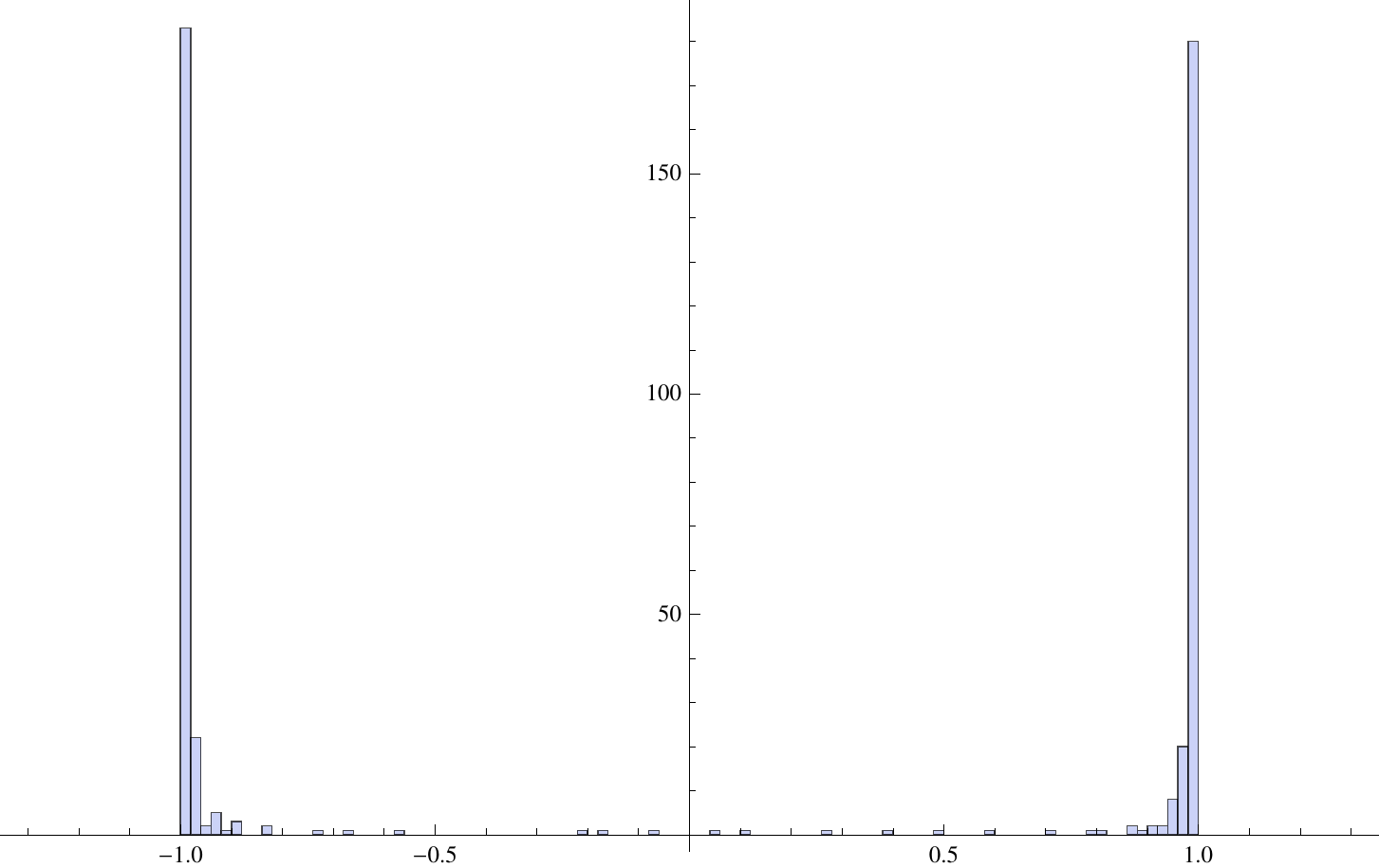}
\par\end{centering}
\caption{{\footnotesize Histogram of the values of $y_{\nu i}$ coordinates showing that most of them are on hypercube vertices.}}
\label{fig:dispersionWR1}
\end{figure}

Our aim was to obtain both statistically independent and binary data. To check this last properties is easier since we have just to check if the patterns $\vec{y}_\nu$ rest on hypercube vertices. This can be seen from Figure~\ref{fig:dispersionWR1} that shows an histogram of the values of coordinates $y_{\nu i}$ (obtained minimizing $U_H$) that shows that this is true as expected.

To verify the independence of data (\ref{eq:y_independance}) with the reduced statistics of this simulation is a challenging task. A necessary condition is that the marginal distributions $p\left(y_{i}=\pm1\right)=\frac{1}{2}$ i.e.\ that each neuron cuts the input data set $\{ \vec{x}_\nu \}$ exactly in $2$ parts. In our simulation this is perfectly achieved, since we got $\frac{m\times h}{2}=\frac{7\times64}{2}=224$ positive coordinates, and $224$ negative ones. Moreover each of the $h=64$ output neurons has for the $m = 7$ inputs exactly $3$ positive and $4$ negative coordinates (or viceversa) suggesting that if we had a larger (and even) number of initial examples, we would get that each neuron would have $m/2$ positive and negative coordinates.

\bigskip

\begin{figure}[h]
\noindent \begin{centering}
(a)\includegraphics[scale=0.25]{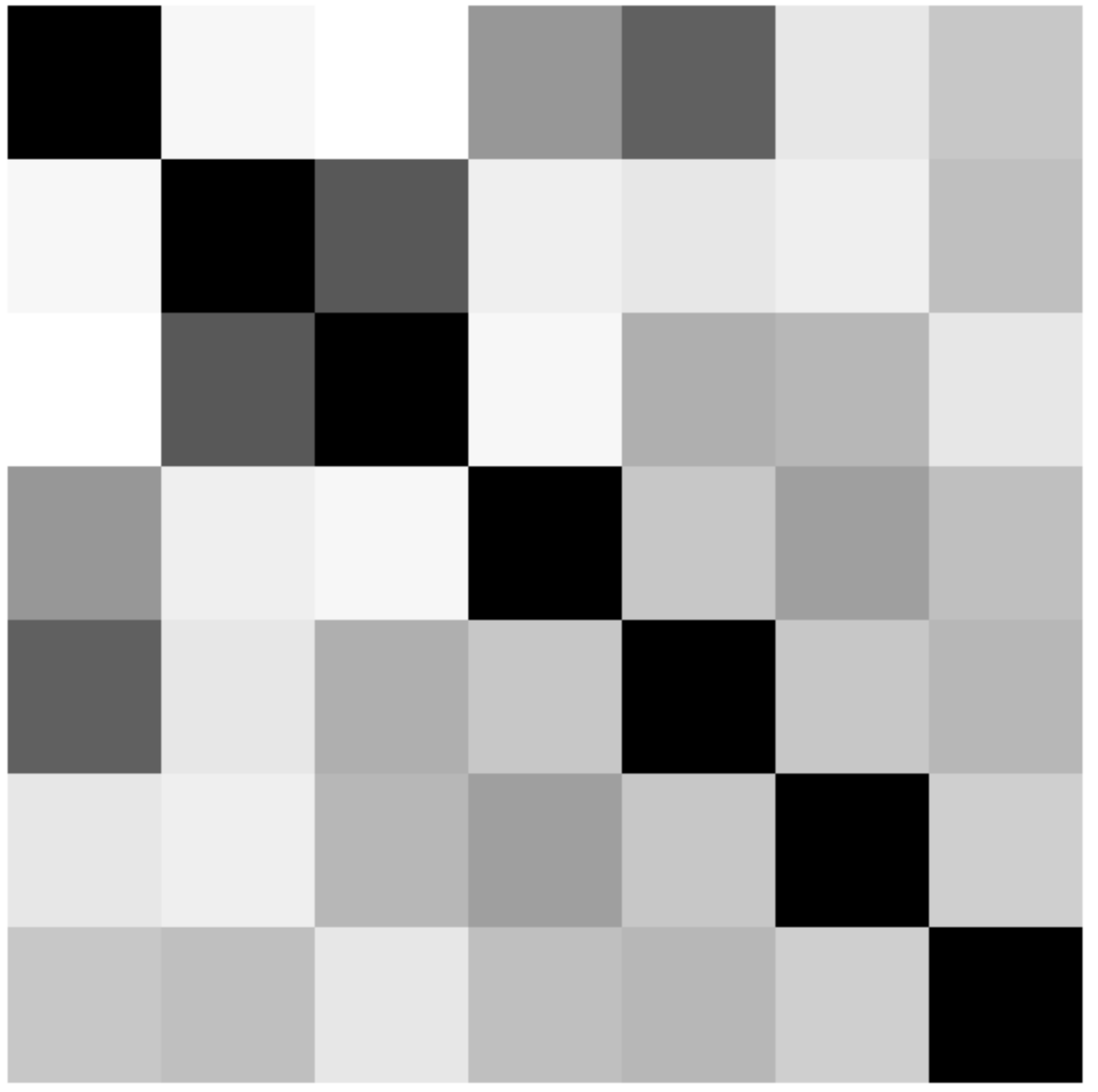}
(b)\includegraphics[scale=0.25]{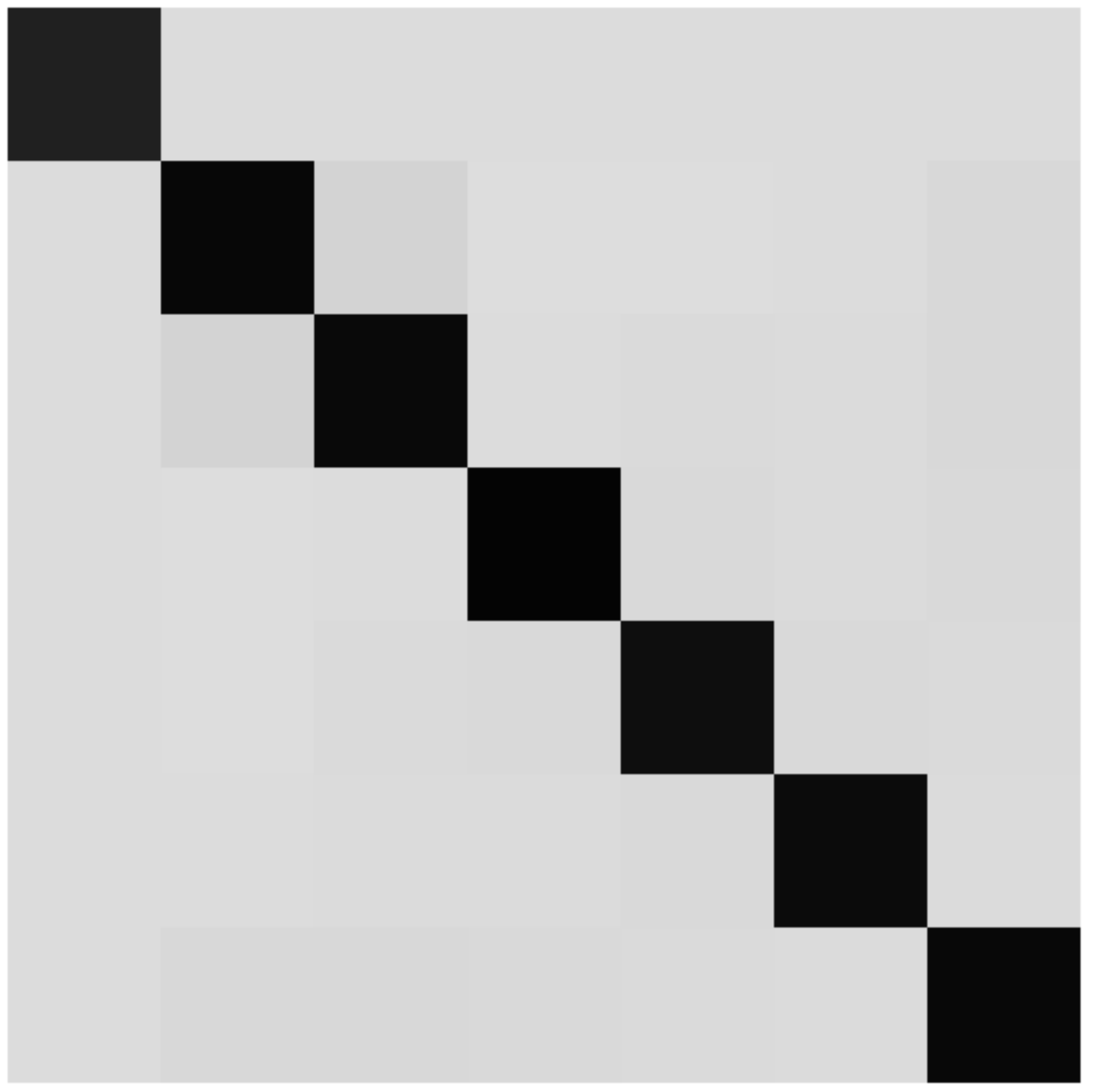}
\par\end{centering}
\caption{{\footnotesize Matrices of scalar products $\vec{y}_\nu \cdot \vec{y}_\mu$ (converted to grayscale) for the systems defined by the Euclidean (a) and the Hamming (b) distance; the outputs are substantially equally spaced in both cases. From a computational point of view it turned out that the NR version that made use of the Hamming distance converged faster than the other: this may suggest, as expected, that it succeeds in providing a greater gradient.}}
\label{fig:correlations}
\end{figure}

To investigate the quality of the solution we analyzed the relative distances of the output data $\vec{y}_\nu$ since one can expect, once (\ref{eq:simplified_potential}) has been minimized, that all relative distances should be equal indicating a roughly constant hypersurface charge distribution. We did this calculating the $m\times m$ matrix of elements $\vec{y}_\nu \cdot \vec{y}_\mu$, that, when $\vec{y}_\nu$ sit on hypercube vertices, represents substantially the distance. In order to make it easier to understand we converted these values to a grayscale ($-h \to$ white, $h \to$ black) and the result is shown in Figure~\ref{fig:correlations}. We can conclude that the $m$ outputs are substantially equally spaced particularly in the second case.

%

\section{\label{sec:Conclusions}Conclusions}
We presented a new approach to the problem of data preprocessing by a layer of Perceptrons: we treat each data vector as a point-like electric charge confined in a $h$-dimensional hypercube, subject to simple Coulomb repulsive forces. We then let the system evolve as it were a real physical system, that is, until it reaches the minimum of the electrostatic energy. At this point, we expect that the charges will occupy the hypercube's vertices and will be as far as possible from each other.

The potential energy function to minimize is continuos (since such is the transfer function $\tanh(x)$), well shaped and, as far as we know, without the relative minima that plague so many cases in neural networks. For these reasons in this case it's sensible to implement a simple gradient descent that produces a strictly local learning rule that is very similar to a Hebb rule with the difference that to update a synapse one needs \emph{all the data} and not just the last seen one.

In our tests this learning algorithm doesn't shine for its speed but one can speculate that for actual calculations one could use more refined minimization of the potential $U$ exploiting the relaxation techniques used routinely for similar electrostatic problems.

Even with a continuous transfer function at the end one obtains binary and statistically independent data that in turn guarantee that the entropy of the output is maximized.

Another characteristics of this network is that one can freely choose the number $h$ of output neurons without any adjustment of the learning algorithm. For small values of $h$ the network implements compression of the incoming data, for larger $h$ just a dimensional reduction without any information loss. For even larger values of $h$ one introduces redundancy in the data useful for subsequent error correction.

Despite some encouraging results we feel that there still is ample space for further theoretical and computational developments.


\bibliographystyle{plain} 
\bibliography{mbh}

\end{document}